\documentclass[prd,11pt]{revtex4-1}

\usepackage{bm}
\usepackage{amsmath}    
\usepackage{graphicx}   
\usepackage{verbatim}   
\usepackage{color}      
\usepackage{subfigure}  
\usepackage{hyperref}   
\definecolor{amaranth}{rgb}{0.9, 0.17, 0.31}
\definecolor{purple(munsell)}{rgb}{0.62, 0.0, 0.77}
\definecolor{americanrose}{rgb}{1.0, 0.01, 0.24}
\definecolor{palatinateblue}{rgb}{0.15, 0.23, 0.89}
\definecolor{royalblue(web)}{rgb}{0.25, 0.41, 0.88}
\definecolor{hanpurple}{rgb}{0.32, 0.09, 0.98}
\definecolor{beaublue}{rgb}{0.74, 0.83, 0.9}
\definecolor{carminered}{rgb}{1.0, 0.0, 0.22}
\definecolor{brightpink}{rgb}{1.0, 0.0, 0.5}
\definecolor{vividviolet}{rgb}{0.62, 0.0, 1.0}

\hypersetup{ linktoc=all,
    colorlinks, linkcolor={palatinateblue},
    citecolor={brightpink}, urlcolor={amaranth}}

\newcommand{\be}{\begin{equation}}
\newcommand{\ee}{\end{equation}}
\newcommand{\bs}{\begin{split}} 
\newcommand{\bea}{\begin{eqnarray}}
\newcommand{\eea}{\end{eqnarray}}

\begin{document}

\title{  DSR, Optics and Electrodynamics } 

\author{Nosratollah Jafari}\email{nosrat.jafari@fai.kz}

\affiliation{Fesenkov Astrophysical Institute, 050020, Almaty, Kazakhstan;
\\
Al-Farabi Kazakh National University, Al-Farabi av. 71, 050040 Almaty, Kazakhstan;
\\
 Center for Theoretical Physics, Khazar University, 41 Mehseti Street, Baku, AZ1096, Azerbaijan}

\begin{abstract}
 
 We investigate some interesting solutions in the DSR theories. These solutions have important features in optics and mechanics. We use these similarities for a better understanding of these theories. We know that, the vacuum in the DSR has an effective refractive index which depends on the frequency of the light. We find this modified refractive index which leads us to a modified Electrodynamics with some new interesting aspects, for example near the Planck energy the electric field can be stronger or
weaker than the usual case. Also, we will find a new modified Lorentz transformations in the first order of the Planck length  which are very similar to the usual Lorentz transformations.
\end{abstract}

\maketitle

\tableofcontents

\section{Introduction}\label{intro}
Proposals of the DSR theories \cite{Amelino-Camelia:2000stu, Amelino-Camelia:2000cpa, Magueijo:2001cr} and finding solutions \cite{Bruno:2001mw, Wang:2013bta, Bruno:2002wc, Jafari:2020ywd}
for them have more than two decades history. DSR theories are nonlinear deformations \cite{Kimberly:2003hp, Salesi:2016ads, Abanto:2021vpr},
which contain two coupled nonlinear hyperbolic differential equations \cite{Jafari:2020ywd}. 

Solutions of the DSR have important features and are similar to the  optics and mechanics equations. We use these similarities for a better understanding of the aspects of the DSR theories. In the DSR theories, the speed of light is different from the special relativistic value c 
\cite{Amelino-Camelia:1999hpv}. In other words, vacuum in the DSR theories has an effective refractive index which depends on the frequency (or energy) of the light \cite{Ellis:2008gg}   

In this paper we investigate further the  solutions of the nonlinear hyperbolic differential equations which we have found in \cite{Jafari:2020ywd} for the DSR theories, and we show that some specific cases of these solutions have interesting optical and mechanical properties. They are related to the Rayleigh and the Lienard equations in optics.

Maxwell equations have been studied in the context of the DSR previously \cite{Takka:2019him}, and in the context of the $\kappa$-Minkowski \cite{Harikumar:2010gq}. But, they have not obtained an explicit form for the modified Maxwell's equations. In this paper we will find the explicit forms of the modified Maxwell's equations in the context of the DSR. 

Also, we will find a new modified Lorentz transformations which are very similar to the usual Lorentz transformations, and the modifications are in the first order of the Planck length. These transformations are very simple and can be useful in the DSR literature. In fact, it can be a simple example of DSR similar to  the MS DSR \cite{Magueijo:2001cr}.

\section{DSR theories at the first order of the Planck length }

The generator of the DSR theories in (3+1) dimensions is given by \cite{Majid:1994cy,Amelino-Camelia:2002ymw,Kowalski-Glikman:2002eyl},
              \be \label{ModGen}  N_1= p_i \frac{\partial  }{ \partial p_0 } + \Big(\frac{l_p}{2} \textbf{p}^2 +
              \frac{1 - e^{-2l_p p_0} }{2l_p} \Big)\frac{\partial  }{ \partial p_i } - l_p p_i  \Big( p_j\frac{\partial }{ \partial p_j } \Big).   \ee
In the leading order of $l_p$, we modify this generator a little bit to be more general as 
 \be \label{3DimGenerator} \Tilde{N}_1=   (p_i +  l_p\beta_0 p_0p_i ) \frac{\partial  }{ \partial p_0 } + 
 \Big( p_0 +  l_p\beta_1 p_0^2+ l_p\beta_2 \textbf{p}^2 \Big)\frac{\partial  }{ \partial p_i } + 
 l_p\beta_3p_i \Big( p_j \frac{\partial  }{ \partial p_j } \Big)  + l_p\beta_4 \epsilon_{1jk} p_j \frac{\partial  }{ \partial p_k },\ee
 where $\beta_0$, ..., $\beta_4$ are arbitrary real numbers. 
 
Thus, we can find differential equations which govern the evolution of energy and momentums in momentum space \cite{Jafari:2020ywd}. For the $p_0$ and $p_i$ components the second order differential equations are
\bea  \label{SecOrderDEBoost}  \left\{\begin{array}{lr}  \frac{d^2p_0}{d\xi^2} = p_0 + l_p ( \beta_0+ \beta_2 + \beta_3)(\frac{dp_0}{d\xi})^2 +
l_p (\beta_0+ \beta_1 )p_0^2 + l_p  \beta_2 (p_2^2 +p_3^2),
\\\\
  \frac{d^2p_1}{d\xi^2} =  p_1 + l_p( \beta_0 +2\beta_1+ 2\beta_2 +  2\beta_3 )p_1\frac{dp_1}{d\xi}. \end{array} \right. \eea

For the $ p_2$ and $p_3$ components the differential equations are
\be  \label{p2p3_DifEq} \left\{\begin{array}{cl}  
    \frac{d p_2}{d \xi}  &=  l_p\beta_3p_ip_2 - l_p\beta_4 p_0p_3, \\\\
    \frac{d p_3}{d \xi}  &=  l_p\beta_3p_ip_3 + l_p\beta_4 p_0p_2.  \end{array}\right.\ee 
Solutions of the differential equations in Eq.~(\ref{SecOrderDEBoost}) and Eq.~(\ref{p2p3_DifEq}) give us the finite-boost DSR transformations for all orders of the rapidity parameter $\xi$, but only in the first order of the Planck length \cite{Jafari:2020ywd}.

  \section{Speed of light, refractive index and modified Maxwell equations at the first order}    
  
  The dispersion relation at the first order of the Planck length will be \cite{Jafari:2020ywd}

 \be   p_0^2- \textbf{p}^2 - 2 l_p \Big(\frac{\beta_0 - \beta_1 + 2\beta_2 +2 \beta_3 }{3} \Big)p_0^3 
 + 2 l_p (\beta_2+ \beta_3) p_0 \textbf{p}^2 =m^2. \ee
 
  By using this dispersion relation the velocity of the photons will be 
\be \label{Vlinear1}  v\simeq  c \Big[1 + \Big( \frac{\beta_0 -\beta_1 - \beta_2 -\beta_3 }{3} \Big) l_p E \Big]. \ee 

We take $\alpha_1 = (\beta_0 -\beta_1 - \beta_2 -\beta_3)/3  $, then we can write a simpler relation for the speed of ligth 

\be \label{Vlinear2}  v\simeq  c \Big[1 + \alpha_1 l_p E \Big]. \ee 

We  define a modified permittivity $\epsilon_{QG}$ and the permeability $\mu_{QG}$ for quantum vacuum, thus we can express the light speed  as 

\be \label{Vlinear3}  v\simeq \frac{1}{\sqrt{\epsilon_{QG} \mu_{QG} } }. \ee 

The refractive index will be 

\be \label{RefIndex}  n= \frac{c}{v}= \frac{1}{1 + \alpha_1 l_p E } \simeq 1 - \alpha_1 l_p E . \ee 

Also, we know that 

  \be n= \sqrt{  \frac{\mu_{QG}}{\mu_0}  \frac{\epsilon_{QG}}{\epsilon_0} }, \ee
  
  where $ \epsilon_0 $  and  $\mu_0$  are the permittivity  and the  permeability of the vacuum in special relativity. We take 
     $ \mu_{QG}/\mu_0 =  \epsilon_{QG}/\epsilon_0   $  (there are other possibilities but this chioce is the simplest), then expressions for the permittivity  and permeability  of the vacuum in 
     quantum gravity will be
     
     \be  \left\{\begin{array}{cl}  
    \epsilon_{QG} =\epsilon_0 ( 1 - \alpha_1 l_p E ), \\\\
    \mu_{QG}= \mu_0 (  1 - \alpha_1 l_p E ).  \end{array}\right.\ee 
    
    From these modified permittivity  and permeability , we reach to the modified Maxwell equations which are given by
    
    \be  \label{} \left\{\begin{array}{cl} \epsilon_0 \Big( 1 - \alpha_1 l_p E \Big)   \nabla. \textbf{E}_{el} &= \rho , \\
      \nabla \times \textbf{B}  -\mu_0 \epsilon_0 \Big( 1 - \alpha_1 l_p E \Big)^2  \partial \textbf{E}_{el} / \partial t       &= \mu_0\Big( 1 - \alpha_1 l_p E \Big)   \textbf{J},\\
      \nabla \times \textbf{E}_{el}   +\mu_0 \Big( 1 - \alpha_1 l_p E \Big)   \partial \textbf{B} / \partial t   &= 0,\\
           \nabla.\textbf{B}   &= 0.
       \end{array}\right.\ee

    These modified Maxwell equations have interesting properties which do not occur in usual case. For instance, the electric field is modified as 
    
     \be    E_{el}= \frac{1}{4 \pi \epsilon_{QG} }  \frac{Q }{r^2} = (1 + \alpha_1 l_p E) \frac{1}{4 \pi \epsilon_0}
     \frac{Q }{r^2},   \ee
     
 where $ E_{el}$ is the electric field at the distance $r$ from the charge $Q$, and $E$ is the energy of this charge as a particle
 
     \be  E= \sqrt{ m^2 c^2 + p^2 c^4 }, \ee
     
     in which $m$ and $p$ is the mass and momentum of the charge as a particle. It is interesting that depending on the sign of the $\alpha_1$  near to the Planck energy the electric field can be stronger or weaker than the usual case.

\section{A specific boost at the first order: Rayleigh and Lienard equations  }
In Eq.~(\ref{SecOrderDEBoost}) and Eq.~(\ref{p2p3_DifEq})   we take  only the $ \beta_1$ to be non-zero and in the following we name it as $ \beta$, also we put the other $ \beta_i$  equal to zero $ \beta_0=\beta_2 =\beta_3=\beta_4 =0 $.  Thus, the simplified differential equations become

 \be  \label{SecondEQ_firstPlanck} \left\{\begin{array}{cl}   \frac{d^2 p_0}{d \xi^2} &=  p_0 +\frac{\beta }{M_p}p_0^2, \\
      \frac{d^2 p_1}{d \xi^2}  &=  p_1 + \frac{2 \beta }{M_p}p_1 \frac{d p_1}{d \xi},\\
      \frac{d^2 p_2}{d \xi^2}  &= 0,\\
         \frac{d^2 p_3}{d \xi^2}  &= 0.
       \end{array}\right.\ee

      The first equation is the the well-known \textit{Rayleigh} equation, and the second equation is the  \textit{Lienard}  equation.

Solutions of these differential equations will give the  transformations which are

\be \label{} \left\{\begin{array}{cl}    p'_0 &= \Big( 1- \frac{ \beta \eta}{M_p}  \Big) \cosh{(\xi)} p_0+ \Big( 1- \frac{ 2\beta \eta}{M_p}    \Big) \sinh{(\xi)} p_1 + 
 \frac{ \beta \eta}{M_p}  \Big[ \cosh(2\xi) p_0 + \sinh(2\xi) p_1 \Big]
\\  \\
\label{} p'_1 & = \Big( 1- \frac{ \beta \eta}{M_p}   \Big) \sinh{(\xi)} p_0+ \Big( 1- \frac{ \beta \eta}{M_p}   \Big) \cosh{(\xi)} p_1 
+  \frac{ 2 \beta \eta}{M_p}   \Big[ \sinh(2\xi) p_0 + \cosh(2\xi) p_1 \Big],\\
p'_2  &= p_2,\\
  p'_3 & = p_3,
\end{array}\right. \ee

These new modified Lorentz transformations are very similar to the usual Lorentz transformations and they only have been modified in the first order of the inversse of Planck energy $ 1/M_p  $.

\acknowledgments 

This research is funded by the Science Committee of the Ministry of Science and Higher Education of the Republic 
of Kazakhstan Program No. BR21881880.

\bibliography{main}

\end{document}